\documentclass[aps,prl,reprint,twocolumn,superscriptaddress,english]{revtex4}

\usepackage{amssymb}
\usepackage{graphicx}
\usepackage{amsmath}
\usepackage[latin9]{inputenc}
\usepackage{array}
\usepackage{multirow}
\usepackage{color}
\usepackage{esint}
\usepackage{bm}
\usepackage{color}
\usepackage{bbm}
\usepackage{hyperref}
\usepackage{babel}
\usepackage{titlesec}
\usepackage{graphicx,amssymb,color}
\usepackage{epsfig}
\usepackage{epstopdf}
\usepackage{mathtools}
\usepackage{soul}
\usepackage[dvipsnames]{xcolor}
\usepackage{tikz}

\begin{document}

\title{ Collective excitations in two-dimensional SU($N$) Fermi gases with tunable spin} 

\author{Chengdong He}
\affiliation{Department of Physics, The Hong Kong University of Science and Technology,\\ Clear Water Bay, Kowloon, Hong Kong, China}
\author{Zejian Ren}
\affiliation{Department of Physics, The Hong Kong University of Science and Technology,\\ Clear Water Bay, Kowloon, Hong Kong, China}
\author{Bo Song}
\affiliation{Department of Physics, The Hong Kong University of Science and Technology,\\ Clear Water Bay, Kowloon, Hong Kong, China}
\author{Entong Zhao}
\affiliation{Department of Physics, The Hong Kong University of Science and Technology,\\ Clear Water Bay, Kowloon, Hong Kong, China} 

\author{ Jeongwon Lee}
\affiliation{Institute of Advanced Study, The Hong Kong University of Science and Technology,\\ Clear Water Bay, Kowloon, Hong Kong, China}
\author{Yi-Cai Zhang}
\affiliation{School of Physics and Electronic Engineering, Guangzhou University, Guangzhou, China}
\author{Shizhong Zhang}
\affiliation{Department of Physics and Center of Theoretical and Computational Physics
,\\ University of Hong Kong, Hong Kong, China}
\author{Gyu-Boong Jo}
\email{gbjo@ust.hk}
\affiliation{Department of Physics, The Hong Kong University of Science and Technology,\\ Clear Water Bay, Kowloon, Hong Kong, China}

\date{\today}
\begin{abstract}
We measure collective excitations of a harmonically trapped two-dimensional (2D) SU($N$) Fermi gas of $^{173}$Yb confined to a stack of layers formed by a one-dimensional optical lattice. Quadrupole and breathing modes are excited and monitored in the collisionless regime $\lvert\ln(k_F a_{2D})\rvert\gg 1$ with tunable spin. We observe that the quadrupole mode frequency decreases with increasing number of spin components due to the amplification of the interaction effect by $N$ in agreement with a theoretical prediction based on 2D kinetic equations. The breathing mode frequency, however, is measured to be twice the dipole oscillation frequency regardless of $N$. We also follow the evolution of collective excitations in the dimensional crossover from two to three dimensions and characterize the damping rate of quadrupole and breathing modes for tunable SU($N$) fermions, both of which reveal the enhanced inter-particle collisions for larger spin. Our result paves the way to investigate the collective property of 2D SU($N$) Fermi liquid with enlarged spin.
\end{abstract}

\maketitle

Recent advances in ultracold alkaline earth-like atoms~\cite{He:2019bb} have opened new possibilities to investigate large spin physics in fermionic systems with SU($N$) symmetry~\cite{Cazalilla:2014kq}. The strong decoupling of nuclear spin ($I$) from electronic angular momentum of these atoms leads to SU($N$) symmetric interactions with $N=1,\cdots, 2I+1$ tunable by controlling their nuclear spins. There have been growing interests in utilizing the enlarged spin symmetry to simulate various quantum phenomena ranging from SU(3) symmetric quantum chromodynamics~\cite{PhysRevLett.98.160405} to unconventional magnetisms~\cite{PhysRevB.37.3774,Honerkamp:2004hq,PhysRevB.70.094521,PhysRevLett.99.130406}. In addition, enhanced degeneracy arising from the spin symmetry is expected to result in topological order, which is analogous to the quantum Hall effects in multi-valley semiconductors~\cite{PhysRevB.59.13147,PhysRevB.67.125314}. To date, however, the on-going efforts on the experimental realization of SU($N$) degenerate quantum gases have been focused on one-dimensional wires~\cite{Pagano:2014hy}, and three-dimensional (3D) optical lattices~\cite{Taie:2012tb,Hofrichter:2016iq}. While there have been spectroscopic measurements of SU($N$)-symmetric interactions~\cite{Zhang2014}, signatures of higher spin symmetry in the context of collective properties of atoms have not been identified in 2D settings.

Owing to the enhancement of quantum fluctuations, novel features of 2D fermionic systems with spin-$1/2$ have been widely studied such as the high mobility electrons in graphene and high temperature superconductivity in cuprates. Ultracold Fermi gases in an oblate optical trap generate a versatile platform to probe 2D physics by freezing out the motional degrees of freedom along the tightly confined direction~\cite{Jasper}. Since the early observations of 2D Fermi gas~\cite{Turlapov:2010kt,Dyke:2011}, most studies have so far focused on the two-component Fermi gas in 2D including the many-body pairing gap~\cite{Feld:2011gf}, the evolution of pairing along dimensional crossover~\cite{Sommer:2012bs} and the spin transport~\cite{Koschorreck:2013dj,Luciuk:2017iw}. Multi-component fermions with higher spin symmetry can dramatically change the pairing mechanism, which have been discussed in recent theoretical studies~\cite{PhysRevLett.99.130406,PhysRevA.82.063615,PhysRevA.83.063607}. Despite the emerging interest in the role of enlarged spin symmetry in 2D, the experimental realization has remained unexplored in fermionic systems.


In this Letter, we realize a stack of 2D SU($N$) Fermi gases in a one-dimensional (1D) optical lattice, and investigate the effect of spin multiplicity on the collective excitations with fixed atom number per spin and scattering length.  We find that the higher spin multiplicity reduces the oscillation frequency of quadrupole modes. The observed collective mode is in good agreement with numerical calculations using kinetic equations in 2D, which provides an experimental confirmation of enhanced interaction effect in weakly interacting SU($N$) fermions. Furthermore, we explore a dimensional crossover from 2D to 3D for tunable SU($N$) systems. Lastly, we experimentally characterize the damping rate of the quadrupole mode for different SU($N$) fermions. Our work provides an atomic 2D platform with SU($N$) symmetry opening a new possibility to study an unconventional Fermi liquid.


%
%

The experiment begins with a laser cooled Fermi gas of $^{173}$Yb loaded into a crossed optical dipole trap (ODT)~\cite{SongApb16}. The crossed ODT is formed with a 1064~nm laser beam along one axis and counter-propagating 532~nm beams with two separately controllable polarizations along the other axis. During the loading and evaporative cooling in the ODT, the polarizations of the 532~nm beams are perpendicular to each other to suppress interference effects. We create a SU($N\leq6$) Fermi gas with tunable spin by optically blasting the unwanted spin components during the evaporative cooling process~\cite{Song:2016ep}. Since we start from equal spin distribution, the blasting method automatically achieves fixed number of atoms per each spin state. The procedure results in a tunable spin mixture (e.g. $N$=1,...,6) with atom number $N_0\simeq10^{4}$ per spin state at the temperature of $T^{3D}/T^{3D}_F=0.2(1)$, where the Fermi temperature is $T^{3D}_F\simeq150$~nK in a 3D trap with trap frequencies of $(\omega_x, \omega_y, \omega_z)=2\pi\times(120,120,30)$~Hz.

\begin{figure} [tbp]
\includegraphics[width=8.9cm]{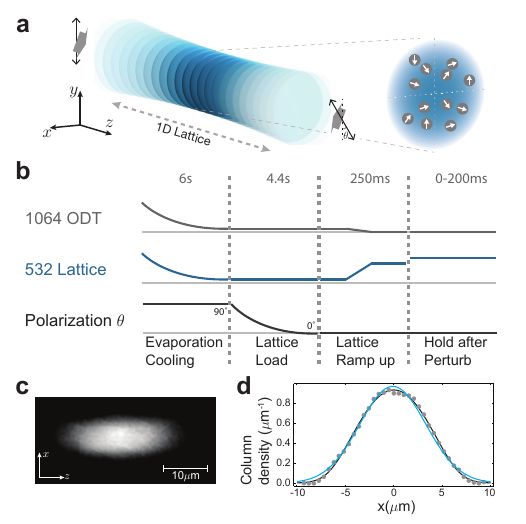}
\caption{\textbf{An array of two-dimensional SU($N$) Fermi gases}. (a) An array of 2D pancake traps is realized by a 1D optical lattice potential formed by counter-propagating 532~nm beams with a tunable relative polarization. (b) During the evaporation cooling process, the polarization angle $\theta$ between incident and reflected beams is $90\textdegree$, leading to zero lattice depth. Subsequently, atoms are loaded into the lattice adiabatically within 4.4~s right after evaporation cooling. By ramping up 532nm laser power, atoms are kept by the 1D lattice only. Also a high lattice depth strongly suppresses  tunneling. Both dipole and quadrupole modes can be excited by a perturbation. (c) An {\it in-situ} absorption image is shown. (d) We measure {\it in-situ} temperature of 2D fermions by obtaining the atomic density $n(x)$ integrated over $y$ direction near the center of the trap. We obtain $T/T^{2D}_F$ from the Fermi-Dirac distribution (grey curve) in contrast to the Gaussian distribution (blue curve). }
	\label{fig1}
\end{figure}

After the preparation of a 3D degenerate Fermi gas, we adjust the polarizations of the 532~nm ODT beams parallel to each other and create an array of 2D traps as shown in Fig.~\ref{fig1}(a). For the purpose, we use a rotational waveplate, allowing us to gradually tune the dimensionality from 3D to 2D with an optical lattice potential of $V(z)=V_0\sin^2(2\pi z/d)$, where $d=$532~nm. The lattice depth is calibrated by the lattice modulation spectroscopy. The lattice loading process consists of two steps (see Fig.\ref{fig1}(b)). First, we rotate the relative polarization of counter-propagating 532~nm beams from perpendicular to parallel over 4.4~s resulting in the lattice depth of $V_0\sim$ 5$E_r$, where $E_r=h\times$4.08~kHz is the recoil energy and  $h$ is the Planck constant. Additional confinement along the $z$ direction is applied to minimize the number of pancakes populated during the lattice loading process, which improved trap homogeneity. Next, we ramp up the lattice depth to the final value of $V_0=53E_r$ within 250~ms, during which the axial confinement along the $z$ direction is switched off. Finally, the sample is adiabatically loaded into the lattice with minimal heating. At this lattice depth, each pancake trap is independent with negligible tunneling energy $J\simeq h\times$0.09~Hz. In each pancake, trapping frequencies are $(\omega_x,\omega_y,\omega_z)\simeq2\pi\times(185,185,59000)$~Hz.  The measured anisotropy $\epsilon=|\omega_x-\omega_y|/2\omega_r$ was less than 0.01, where the radial trapping frequency is defined as $\omega_r=\sqrt{\omega_x\omega_y}\simeq2\pi\times185$ Hz.

We initially investigate the property of the 2D gas by comparing the temperature with the axial confinement of the lattice potential. We measure {\it in-situ} temperature of the Fermi gas in the lattice by fitting to the column density $n(x)$ (see Fig.\ref{fig1}(c,d)). The column density $n(x)=-\frac{\sqrt{m}}{\sqrt{2\pi}\hbar^2\beta^{3/2}\omega_r}\text{Li}_{3/2}(-ze^{-\beta\frac{m}{2}\omega_r^2x^2})$ and the fugacity $z=e^{\beta\mu}$ are related to the temperature by $T/T^{2D}_F=1/\sqrt{-2\text{Li}_2(-z)}$ where $\text{Li}$ is the PolyLog function. In the lattice, we obtain $T\simeq$ 60nK or $T/T^{2D}_F\simeq 0.42$, where $k_B T^{2D}_F = \hbar\omega_r\sqrt{2N_{2d}}$ and $N_{2d}\simeq$100 is the number of atoms per spin in each pancake near the center of the trap. As the condition $E_F,k_B T\ll\hbar\omega_z$ is fulfilled, majority of the atoms occupy only the ground level of the harmonic oscillator.

Our main result is the observation of a change in quadrupole mode of the 2D Fermi gas with spin multiplicity $N=1,\cdots,6$. Collective modes of trapped fermions have been widely used to reveal interaction effects, as shown in experiments with two-component Fermi gases in 2D~\cite{Vogt:2012dm,PhysRevLett.121.120401,Peppler:2018fb}. In 2D, an interaction parameter is given by $g_{2D}=g_{3D}(\sqrt{2\pi}l_z+a_{3D}\ln(B\hbar\omega_z/2\pi E_F))^{-1}$ with $B=0.915$~\cite{Petrov:2001}. Here $l_z=\sqrt{\hbar/m\omega_z}$ is a harmonic oscillator length along the tightly confined direction and $g_{3D}=4\pi \hbar^2a_{3D}/m$ is a 3D interaction parameter with the $s$-wave scattering length $a_{3D}$. Correspondingly, the 2D scattering length is given by $a_{2D}=\sqrt{\pi/B}l_z\exp(-\sqrt{\pi/2}l_z/a_{3D})$~\cite{Jasper}. For our experimental parameters, the collective excitations are well described in the collisionless regime with the dimensionless parameter $\lvert \ln (k_Fa_{2D})\rvert\gg 1$. For a two-component Fermi gas at $T=0$, this leads to a shift of quadruple frequency as $\omega_q/\omega_d=\sqrt{2(2+\chi)/(1+\chi)}\simeq (2-\chi/2)$~\cite{Ghosh:2002fs}, where $\chi=-\ln^{-1}(k_Fa_{2D})=g_{2D}\frac{m}{2\pi\hbar^2}>0$.
\begin{figure} [tbp]
	\includegraphics[width=8.3cm]{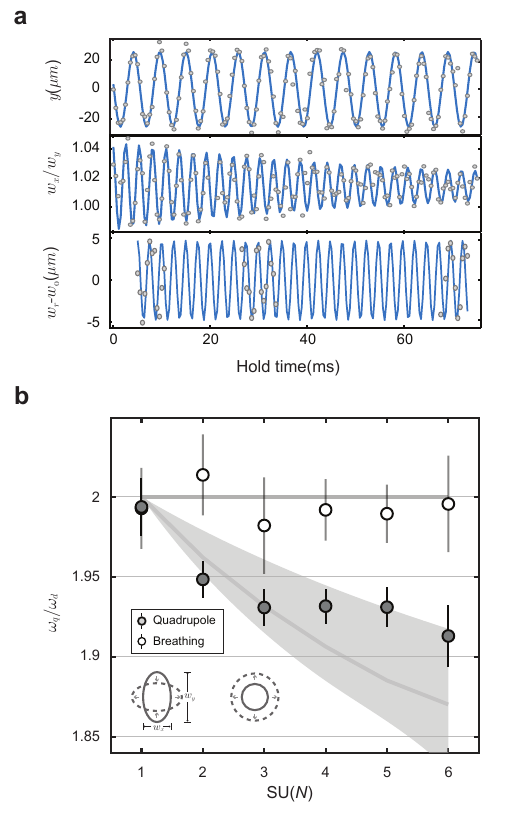}
	\caption{\textbf{Measurement of quadrupole and breathing mode frequencies}. (a)  Oscillations showing dipole (top), quadrupole (mid) and breathing (bottom) modes for a SU(6) gas after the perturbation. To ensure the trap is completely deformed after the sudden increase of the 532~nm lattice power, we monitor the collective oscillations starting from 0.5~ms after the quench. (b) The ratio between quadrupole and dipole modes $ \omega_q/\omega_d$ (solid cirlce) is monitored for different number of spin components $N$ with atom number being fixed per spin in comparison with theoretical prediction (solid curve). The shaded region indicates the uncertainty on the theoretical values resulting from the experimental uncertainty. For comparison, we measure the breathing mode (open circle) by measuring the width of the cloud $w_r=\sqrt{w_x w_y}$ subtracted by the equilibrium width $w_0$. The error bar includes statistical and systematic errors of measurements.}
	\label{fig2}
\end{figure}

\begin{figure} 
	\includegraphics[width=7.3cm]{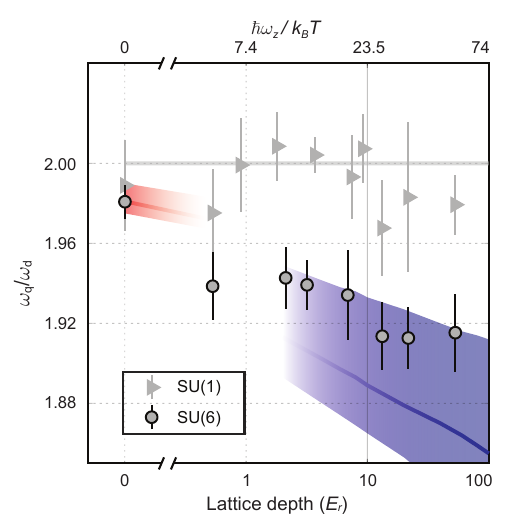}
	\caption{\textbf{Dimensional crossover from 2D to 3D}. We measure the change in quadrupole mode of SU(1) and SU(6) Fermi gases to find the signature of dimensional crossover. The Fermi gas becomes increasingly confined to a stack of 2D layers as we increase the lattice depth of 1D optical lattice. In contrast to non-interacting fermions (triangle, SU(1)) with the quadrupole mode being close to 2$\omega_d$ (grey line), decreasing the dimensionality leads to reduced frequency of quadrupole mode for SU(6) fermions (circle) due to amplification of the interaction effect. The blue and red solid lines indicate theoretical prediction in 2D and 3D, respectively. The shaded region indicates the uncertainty on the theoretical values resulting from the experimental uncertainty. The error bar includes statistical and systematic errors of measurements.}
	\label{fig3}
\end{figure}

To model the collective dynamics, we employ the kinetic equation for semi-classical distribution function $f_{\alpha\beta}({\bf r}, {\bf p})$, where $\alpha,\beta=1,2,\cdots, N$ label the spin components. Assuming no off-diagonal coherence during the collective motion, $f_{\alpha\beta}({\bf r},{\bf p})=f_{\alpha}({\bf r},{\bf p})\delta_{\alpha\beta}$, and taking into account the mean-field terms, the  kinetic equation takes the form
\begin{align}
\frac{\partial f_{\alpha}({\bf r},{\bf p})}{\partial t}+{\bf p}\cdot\nabla_{\bf r} f_\alpha-\nabla_{\bf r}(V+V_{\rm mf})\cdot\nabla_{\bf p}f_\alpha=I_{\rm col}\label{kineticeqn}
\end{align}

where $V_{\rm mf}({\bf r})=g_{2D}\sum_{\beta\neq \alpha} n_\beta({\bf r})$ encapsulates the effects of interaction in 2D, the collisional integral $I_{\rm col}$ and $V(\bf r)$ the external trap. In obtaining solutions to eq.(\ref{kineticeqn}), we use the scaling form for $f_{\alpha}({\bf r},{\bf p})$ as detailed in the Supplementary Material~\cite{supp}. In this formulation, the effect of spin multiplicity enters $V_{\rm mf}({\bf r})$ as a multiplicative constant and consequently, the modification to quadruple mode frequency scales approximately linearly with $N$ as $2\omega_{d}-\omega_q\propto (N-1)g_{2D}$ for our experimental conditions, which corresponds to the collisionless regime ($I_{\rm col}=0$) with $\ln(k_F a_{2D})= -4.3$. This quadrupole frequency shift amplified by spin multiplicity is generic even in 3D.

\begin{figure} 
	\includegraphics[width=8.1cm]{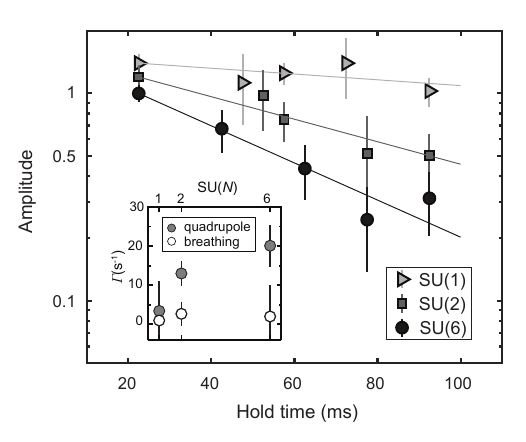}
	\caption{\textbf{Damping of quadrupole and breathing modes for different spin multiplicity $N$}. After inducing collective excitations, we monitor the amplitude of the oscillation up to 100~ms hold time. We extract the damping rate by fitting the oscillation amplitude with an exponential decay curve  with zero amplitude fixed at long hold time. In the inset, we show the damping rates of quadrupole and breathing modes for different spin multiplicity $N$. Here, the error bar indicates the fit error.}
	\label{fig4}
\end{figure}

To induce collective excitations in our experiment, we abruptly increase the radial trap frequency by 10$\%$  as shown in Fig.~\ref{fig1}(b). The sudden increase of the lattice depth induces multiple collective modes (e.g. dipole, breathing and quadrupole modes) simultaneously due to the change of gravitational sag during the excitation. Subsequently, collective oscillations are monitored up to 150~ms by turning off the trap at different times followed by a 8~ms time-of-flight expansion. Absorption images of the atomic cloud are taken by the resonant 399~nm $^1$S$_0$-$^1$P$_1$ optical transition.

 In order to identify different collective modes induced simultaneously, we determine the center-of-mass position and the width of the cloud in the $x$ and $y$ direction, $w_x$ and $w_y$, respectively. We calibrate the dipole frequency $\omega_d$ from the center-of-mass position in the $y$ direction (Fig.~\ref{fig2}(a)). The quadrupole and breathing mode frequencies are then obtained by taking $w_x/w_y$ and $\sqrt{w_x w_y}$, respectively (Fig.~\ref{fig2}(a)). The observed quadrupole modes (see Fig. \ref{fig2}(a)) show oscillations near twice the frequency of the dipole mode~\cite{freq} with more pronounced decay rate, which will be discussed later. We measure an extremely small decay rate of the dipole mode frequency $\Gamma_d/\omega_d<0.0006$ due to the minimal trap anharmonicity reflected in small trap anisotropy $\epsilon$. Consequently, the dipole mode is  precisely determined within 0.5 $\%$ or $\omega_d$=$2\pi\times$185(1)~Hz.

 In Fig.~\ref{fig2}(b), we plot the ratio of $\omega_q/\omega_d$, which provides direct access to the interaction effect in SU($N$) gases, for various spin multiplicities. We find a clear deviation of $\omega_q/\omega_d$ away from 2 as the spin multiplicity increases, consistent with the theoretical expectation based on kinetic theory, $2\omega_{d}-\omega_q\propto (N-1)g_{2D}$ which only takes into account mean field effects. The theoretical curve shown as a solid line is based on numerical solutions of the kinetic equation (\ref{kineticeqn}) in the mean-field approximation and are in reasonable agreement with observations considering experimental uncertainty.  We also measure the breathing mode by geometric averaging of $w_x$ and $w_y$. In contrast to the quadrupole mode, the breathing mode does not depend on spin multiplicity being consistent with the classical scale invariance in a weakly interacting gas~\cite{Pit1997,Vogt:2012dm}; while quantum anomaly, the breakdown of classical scale invariance, has recently been observed in the strongly interacting regime~\cite{PhysRevLett.121.120401,Peppler:2018fb}.

 In Fig.~\ref{fig3}, we investigate the SU($N$) dependent quadrupole mode frequencies along the dimensional crossover from 2D to 3D, by controlling the lattice depth $V_0$. In the 2D limit where the inter-layer coupling is not taken into account, the quadrupole mode frequency can be estimated in the mean-field regime as a function of the lattice depth, $\omega_q \propto 2\omega_{d}-(N-1)g_{2D}$ where $g_{2D}\propto V_0^{1/4}$ ignoring the term with $a_{3D}$, as indicated by the blue region in Fig.~\ref{fig3}. The observed quadrupole frequency is reasonably consistent with 2D prediction even in the intermediate lattice depth where the Fermi energy is comparable to the lattice depth. In 3D limit, however, we observe that $\omega_q/\omega_d$ approaches 2 as the trapping geometry becomes closer to the 3D regime of $k_B T\gg\hbar\omega_z$ due to the small interaction parameter $g_{3D}$ as shown in Fig.~\ref{fig3} (red shaded region). As a reference, $\omega_q/\omega_d$ is also monitored for the non-interacting spin-polarized Fermi gas (i.e. SU(1)), which remains around 2 throughout the same range of trapping parameters. Our results highlight the role of lower dimensions for amplification of negligibly small interaction effects. To fully calculate the collective mode in the crossover regime, however, the inter-layer coupling needs to be further considered.


Finally, we turn our attention to the damping process of the collective excitation in a 2D SU($N$) Fermi gas. Fig.~\ref{fig4} shows the evolution of the quadrupole and breathing mode amplitude for SU(1), SU(2) and SU(6) gases at the lattice depth of 53$E_r$. As spin multiplicity increases, quadrupole oscillations exhibit more obvious damping effect. This phenomenon can be explained by noticing that in a SU($N$) Fermi gas, the relaxation of the quadrupole mode is determined by the appropriate damping of moment $\chi=\sum_i(x_{i,\sigma}^2-y_{i,\sigma}^2)$ with $x_{i\sigma}$ and $y_{i\sigma}$ gives the position of the particle $i$ with spin-$\sigma$. The rate of damping for $\chi$ is proportional to the collision integral $\langle\chi I_{\rm col}\rangle$~\cite{Vichi2000, Silvia2013}. Within the simplest assumption in which $f_{\alpha}$ is independent of $\alpha$, $I_{\rm col}$ is proportional to ($N-1$) and this leads to larger collision integral and consequently, a faster decay. On the other hand, the breathing mode suffers much less damping as function of spin multiplicity due to vanishingly small bulk viscosity in our system. The damping of collective modes could become a useful tool for the detection of the Kondo scattering and pairing states, if a two orbital system is implemented~\cite{Sundar:2016hg,Zhang:2017jqa}.

 We further note that the implementation of optical Feshbach resonances (OFR) can enhance the atomic interactions~\cite{Enomoto:2008bx,Blatt:2011dr}. Despite the atomic loss and the slightly broken SU($N$) symmetry induced by the OFR beam, we expect the SU($N$) symmetry can be effectively maintained within tens of ms with minimal nuclear spin relaxation rate~\cite{Gorshkov2010}, during which the collective mode can be investigated. Pushing the SU($N$) symmetric interaction closer to the strongly interacting case of $\lvert\ln(k_F a_{2D})\rvert\simeq 1$ and searching for its effect in collective modes is one possible extension of our work.

In conclusion, we realize a two-dimensional Fermi gas with tunable spin and detect its SU($N$) symmetric interaction effects using collective excitations. Various collective modes are investigated revealing the decrease in the ratio of quadruple to dipole mode frequency with $N$ in good agreement with mean-field prediction, while the ratio of breathing to dipole mode frequency stayed constant. We also follow the evolution of collective modes in the dimensional crossover from 2D to 3D and measure their damping rates in 2D.  Quantum anomaly~\cite{PhysRevLett.121.120401,Peppler:2018fb} in 2D SU($N$) fermions would be an interesting topic for future studies. In addition, possible extensions of our work can be considered in the context of two-orbital system in 2D~\cite{oppong2019}.

{\bf Acknowledgement} This work has been supported by C6005-17G from the Hong Kong Research Grants Council (HKRGC). G.-B. J. also acknowledges the support from the Croucher Foundation and HKRGC through ECS26300014, GRF16300215, GRF16311516, and GRF16305317, GRF16304918 and the Croucher Innovation Award, respectively. S.Z. acknowledges GRF17303215 and Croucher Innovation Award. Y.-C. Z. was supported by the NSFC under grants No. 11874127 and No. 11747079.


\begin{thebibliography}{28}
\bibitem{freq} During the experiment, we excite one of the dipole modes along the $y$ direction. Due to the small anisotropy of the trap, the radial trapping frequency is given as $\omega_r=$~1.005~$\omega_y$.
\end{thebibliography}

\newpage
\clearpage

\renewcommand{\thesection}{M-\arabic{section}}
\setcounter{section}{0}  
\renewcommand{\theequation}{M\arabic{equation}}
\setcounter{equation}{0}  
\renewcommand{\thefigure}{M\arabic{figure}}
\setcounter{figure}{0}  

\section*{\large Supplementary notes}

{\bf Numerical methods using the kinetic euqation}  
To investigate the collective mode of SU$(N)$ gases in a harmonic trap, we make use of the kinetic equation describing the time evolution of the Wigner distribution function:
\begin{align}
f_{\alpha\beta}({\bf r},{\bf p})=\int d{\bf r'}e^{i{\bf p}\cdot{\bf r'}}\left\langle \psi^\dagger_\alpha({\bf r}+\frac{\bf r'}{2})\psi_\beta({\bf r}-\frac{\bf r'}{2})\right\rangle.
\end{align}
Here the field operator $\psi_\alpha({\bf r})$ creates a fermion with spin projection $\alpha$ at position ${\bf r}$. Its time evolution can be determined by the Heisenberg equation of motion for the operator product $\psi^\dagger_\alpha({\bf r}+\frac{\bf r'}{2})\psi_\beta({\bf r}-\frac{\bf r'}{2})$ given the form of a SU($N$) Hamiltonian of the following form.
\begin{align}
H &=H_0+H_{\rm int}\\
H_0 &=\sum_{\alpha=1}^N\int d{\bf r}\psi^\dagger_\alpha({\bf r})\left(-\frac{\hbar^2\nabla^2}{2m}+V({\bf r})\right)\psi_\alpha({\bf r})\\
H_{\rm int} &=\frac{g}{2}\sum_{\alpha\beta}\psi^\dagger_\alpha({\bf r})\psi^\dagger_\beta({\bf r})\psi_\beta({\bf r})\psi_\alpha({\bf r})
\end{align}
where $V({\bf r})=\frac{1}{2}m(\omega_x x^2+\omega_y y^2+\omega_z z^2)$ is the external trapping potential, currently assumed to be three dimensional and anisotropic. In actual experiment $\omega_x=\omega_y\ll\omega_z$. In later calculations, we shall assume that the $z$-direction confinement is large and the system is effectively two-dimensional. In that case, the three-dimensional coupling constant $g$ needs to be replaced with its 2D counterpart $g_{2D}$. The derivation of transport equations, however, remains unchanged going from 3D to 2D.

In our calculation, we shall assume that no off-diagonal elements are generated in the collective oscillations. This is the case for the monopole and quadruple mode that we excite in our experiments, which corresponds to the total density oscillations, as there is no transverse magnetization in the initial preparation of the sample. As a result, we shall assume only the diagonal part of the Wigner function is nonzero. However, we will take into account the forward scattering (Hartree potential) which supplies the additional mean field that affects the equation of motion of $\psi^\dagger_\alpha({\bf r}+\frac{\bf r'}{2})\psi_\beta({\bf r}-\frac{\bf r'}{2})$. With these simplifications, the time-dependent equation for $f_{\alpha\beta}({\bf r},{\bf p})=f_{\alpha}({\bf r},{\bf p})\delta_{\alpha\beta}$ becomes
\begin{align}
\frac{\partial f_{\alpha}({\bf r},{\bf p})}{\partial t}+{\bf p}\cdot\nabla_{\bf r} f_\alpha-\nabla_{\bf r}(V+V_{\rm mf})\cdot\nabla_{\bf p}f_\alpha=I_{\rm col}
\end{align}
where $I_{\rm col}$ is the collision integral. The mean field potential $V_{\rm mf }({\bf r})$ is due to forward scattering and takes the form $V_{\rm mf}({\bf r})=g\sum_{\beta\neq \alpha} n_\beta({\bf r})$.  In our experiment, the parameter $\omega \tau$ that determines whether the system is in collisional or collisionless regime can be estimated as follows. The typical frequency associated with the oscillation is of order $\omega_r=\omega_x=\omega_y$. Since the system is weakly interacting, it is reasonable to estimate the collision time as $\tau^{-1}=n\sigma v_{\rm F}(T_{\rm F}/T)^2$ where the temperature factor accounts for the Pauli exclusion principle restricting the available scattering phase space. Using parameters appropriate to the experimental situation, this gives $\omega\tau\sim 20\gg 1$, indicating that one is in the collisionless regime. In the calculation below, we shall then neglect the collision integral $I_{\rm col}$.

To solve the equation, we make use of the following ansatz solution for the collective mode~\cite{Odelin2002}.
\begin{align}
f_\alpha({\bf r},{\bf p},t)=f^0_\alpha(\frac{r_i}{\lambda_{\alpha i}(t)},\lambda_{\alpha i}(t)p_i-\dot{\lambda}_{\alpha i}(t)r_i)
\end{align}
$\lambda_{\alpha i}(t)$ describes the deformation of density for $\alpha$-th component along $i$-direction. $f^0_\alpha$ gives the equilibrium distribution of the Wigner function. Substitute this amsatz into the kinetic equation, we obtain a set of differential equations relating the $3N$ variables $\lambda_{\alpha i}(t)$. In the case of small oscillation $|\lambda_{\alpha i}(t)-1|\ll 1$ and it is possible to write $\lambda_{\alpha i}(t)=1+\delta\lambda_{\alpha i}(t)$, one finally obtains
\begin{align}\label{ke}
&\frac{d^2 \delta\lambda_{\alpha i}(t)}{dt^2}+4\omega_i^2\delta\lambda_{\alpha i}(t)+\frac{g}{N_\alpha \langle r_i^2\rangle}\times\\\nonumber
&\sum_\beta A_{\alpha i\beta}(2\delta\lambda_{\alpha i}-\sum_j\delta\lambda_{\alpha j})+\sum_\beta B_{\alpha i;\beta j}(\delta\lambda_{\beta j}-\delta\lambda_{\alpha i}))
\end{align}
where we have defined
\begin{align}
A_{\alpha i\beta} &= \int d{\bf r}n_\alpha \frac{\partial n_\beta}{\partial r_j}r_j\\
B_{\alpha i;\beta j} &= \int d{\bf r}\frac{\partial n_\alpha}{\partial r_i}\frac{\partial n_\beta}{\partial r_j}r_ir_j
\end{align}
where $n_\alpha({\bf r})$ is the equilibrium density distribution of the system. Within semi-classical approximation, we have the following expression
\begin{align}
n_\alpha({\bf r})=\int\frac{d{\bf p}}{(2\pi)^d}\left(e^{\frac{1}{k_B T}(\frac{{\bf p}^2}{2m}+V({\bf r})+V_{\rm mf}({\bf r})-\mu)}+1\right)^{-1}\label{density}
\end{align}
where $T$ is the temperature and $\mu$ is the chemical potential, to be determined by requiring that $\int d{\bf r} n_{\alpha}{\bf r}=N_{\alpha}$. If neglect the spin oscillations, then it is possible to assume that
\begin{align}
&\delta\lambda_{\alpha i}=\delta\lambda_{i}\label{sl}\\
&n_\alpha(r)=n(r)\label{sn}
\end{align}
irrespective of the spin components. With this simplification, one can simplify eq.(\ref{ke}) considerably and our numerical calculation is based on the simplification made above.

In our numerical calculation, we assume that the system can be described as a quasi two-dimensional system with $d=2$. The modified 2D coupling constant is given by~\cite{Petrov}
\begin{align}
g_{\rm 2D}=\frac{g_{\rm 3D}}{\sqrt{2\pi}l_z}\left(1+\frac{a_{3D}\ln\frac{0.915\hbar \omega_z}{\pi E_F}}{l_z\sqrt{ 2\pi}}\right)^{-1}
\end{align}
where $g_{3D}=\frac{4\pi\hbar^2 a_{3D}}{m}$ and $l_z=\sqrt{\hbar/(m\omega_z)}$ is length unit determined by $\hat{z}$-axis trap frequency $\omega_z$.  The equation for scaling factor is
\begin{align}\label{eq2d}
\frac{d^2{\delta\lambda}_{j}}{dt^2}+4\omega^2_j\delta\lambda_{j}+\frac{\kappa A_j}{\langle r^2_j\rangle}
[2\delta\lambda_{ j}-\sum_{j'=x,y}\delta\lambda_{ j'}]=0,\\
A_{j }=\int d^2\vec{r} n(r) \frac{\partial n}{\partial r_j} r_j=-\frac{1}{2}\int d^2\vec{r} n^2(r),
\end{align}
where $n(r)$ is density distribution function in equilibrium states.  $\langle r^{2}_{j}\rangle=\int d^2\vec{r} n(r)r^{2}_{j}$, $\kappa=(N-1)g_{2D}$.  $N$ is number of components. The number of particle in each pancake is assumed to be $100$.

Correspondingly, the density distribution in 2D case is given by eq.(\ref{density}) with integration over 2D momentum space $p_x,p_y$. The mean field $V_{\rm mf}({\bf r})=\kappa n({\bf r})$ according to eqs.(\ref{sl}) and (\ref{sn}). From eq.(\ref{eq2d}), we observe that for breathing frequency ($\delta\lambda_x=\delta\lambda_y$), the interaction does not alter the breathing mode frequency, i.e., $\omega_{b}=2\omega_{r}$. When the interactions are very small, the frequency shifts of quadrupole modes is proportional to interaction strength, i.e.,  $\omega_q-2\omega_{r}\propto\kappa=(N-1)g_{2D}$ (see figure.2 in main text).

%

\end{document}